\documentclass[conference]{IEEEtran}

\usepackage{amsmath}
\usepackage{color}
\usepackage{array}
\usepackage{stfloats}
\usepackage{amsthm} 
\usepackage{amssymb}
\usepackage{float}
\usepackage{nccmath}
\usepackage{graphicx}
\usepackage{setspace}
\usepackage{booktabs}
\usepackage{mwe}
\usepackage{cite}
\usepackage{amsmath,amssymb,amsfonts}
\usepackage{graphicx}
\usepackage{textcomp}
\usepackage{kotex}
\usepackage{xcolor}
\usepackage[utf8]{inputenc} 
\usepackage{kotex} 

\usepackage{algorithm}
\usepackage{algorithmic}
\usepackage{grffile}

\begin{document}
\nocite{*}
\title{Spatio-Temporal Attack Course-of-Action (COA) Search Learning for Scalable and Time-Varying Networks}

\author{
\IEEEauthorblockN{Haemin Lee$^\dag$, Seok Bin Son$^\dag$, Won Joon Yun$^\dag$, Joongheon Kim$^\dag$, Soyi Jung$^\ddag$, and Dong Hwa Kim$^\S$}
\IEEEauthorblockA{
$^\dag$Department of Electrical and Computer Engineering, Korea University, Seoul, Republic of Korea\\
$^\ddag$School of Software, Hallym University, Chuncheon, Republic of Korea\\
$^\S$Agency for Defense Development (ADD), Seoul, Republic of Korea\\
\texttt{\{haemin2,lydiasb,ywjoon95,joongheon\}@korea.ac.kr}, 
\texttt{sjung@hallym.ac.kr}, 
\texttt{dhkim@add.re.kr}}
}

\maketitle

\begin{abstract}
One of the key topics in network security research is the autonomous COA (Couse-of-Action) attack search method. Traditional COA attack search methods that passively search for attacks can be difficult, especially as the network gets bigger. To address these issues, new autonomous COA techniques are being developed, and among them, an intelligent spatial algorithm is designed in this paper for efficient operations in scalable networks. On top of the spatial search, a Monte-Carlo (MC)-based temporal approach is additionally considered for taking care of time-varying network behaviors. Therefore, we propose a spatio-temporal attack COA search algorithm for scalable and time-varying networks.
\end{abstract}


\IEEEpeerreviewmaketitle

\section{Introduction}
As the network becomes more complex and expanded, the importance of security for the network has increased~\cite{isj202003saad,isj2021dao}. Especially, security accidents in the network have become more frequent, and its importance has become obviously higher.

In this situation, it is meaningful to identify the attackers' behaviors; and the corresponding research have been widely and actively conducted in the security and privacy literature with the name of \textit{attack course-of-action (COA)}.

In order to design the attack COA search algorithms, many modern artificial intelligent methods can be utilized such as time-varying dynamic optimization and heuristic search~\cite{jsac201806choi,jsac201811dao,tmc201907koo,twc201910choi,twc201912choi,twc202012choi,twc202104choi,tmc2021yi,tvt2021jung,isj2021jung}, deep and distributed learning~\cite{pieee202105park,tmc202106malik,tvt201905shin,isj202103kim,icdcs2020meteriz}, and reinforcement learning with and without multi-agent cooperation~\cite{tvt202106jung,ijcai2019shin,8851270,9682599}. Among these algorithms, deep learning-based approaches are widely and actively used in modern research trends, however, these approaches are not considered in this paper because it works very well if the attack patterns are similar to the training data patterns. Therefore, the training-based algorithms cannot be scalable. In order to  design and implement the scalable attack COA algorithms, a novel spatial algorithm inspired by heuristic search is proposed in this paper. By defining appropriate heuristic functions for attack COA search and path planning, our proposed spatial attack COA search algorithm can work for any kinds of networks. Furthermore, we additionally consider Monte Carlo (MC)-based tree search in order to consider time-varying network behaviors. Therefore, the final form of our proposed attack COA search algorithm is \textit{spatio-temporal}.

The rest of this paper is organized as follows. 
Sec.~\ref{sec:algorithm} introduces our proposed spatio-temporal attack COA search algorithm in scalable and time-varying networks.
Sec.~\ref{sec:results} presents algorithm execution results over two-types of networks.
Sec.~\ref{sec:conclusions} concludes this paper and presents future research directions.

\section{Spatio-Temporal Attack COA Optimization}\label{sec:algorithm}

Our proposed spatio-temporal attack COA optimization algorithm consists of two parts, i.e., (i) spatial intelligent attack COA heuristic for scalable networks (refer to Sec.~\ref{sec:algo1}) and (ii) MC-based tree search for time-varying networks (refer to Sec.~\ref{sec:algo2}), respectively.

\subsection{Spatial Intelligent Attack COA Heuristic for Scalable Networks}\label{sec:algo1}


Our proposed spatial intelligent attack COA heuristic is for computing optimal attack COA routes. Our proposed algorithm walks through the intermediate network nodes from the attack source node to attack target node which can maximize the impacts of attacks. For this purpose, the value of attack should be numerically defined.


For the purpose, our learning heuristic function can be defined as follows,
\begin{equation}
     f(n) = g(n) + h(n)
\end{equation}
where $f(n)$ is defined as the value of attack at current intermediate network node $n$. In addition, $g(n)$ stands for the summation of values of attacks from the attack source node to current intermediate network node $n$. Lastly, $h(h)$ stands for the approximated value of attack from current intermediate network node $n$ to our attack target node.


Here, $g(n)$ is a calculated value which is the sum of the passed edges to arrive at current node, while $h(n)$ is the approximated value for the valuation of target node which can regarded as a heuristic function. To find the optimal path, the heuristic function $h(n)$ needs to be well-defined. In this paper, we used vulnerability information for all nodes using the Common Vulnerability Scoring System (CVSS)~\cite{cvss}.


This CVSS is a framework used to rank the characteristics and severity of vulnerabilities and a exploitable weaknesses. In CVSS, base score reflects the harmfulness of the vulnerability itself, and the exploitability score reflects the feasibility of exploiting that particular vulnerability which has a maximum value of 10. To define the vulnerability score we used exploitability score to weight the significance of the base score considering how feasible it is to use a certain vulnerability~\cite{hu2020automated}.


We first assign the values of source node and target node as 0.01 and 100, respectively. Assuming that if any node with higher vulnerability score will lead to higher probability to reach the target node, we assigned individual vulnerability score to the nodes that exploits a vulnerability using $Score_{vul}$~\eqref{CVSS} to the node with higher vulnerability score, as follows,

\begin{equation}
     Score_{vul} = baseScore \times \frac{exploitabilityScore}{10}.
     \label{CVSS}
\end{equation}

In addition, for the nodes that access a files or execute a code, we assign the value of 1.5 since such actions are critical. For others, we assign 0.


Here, $g(n)$ evaluates the attack value along the path from source node to current node, which is the sum of the edge weights between the visited nodes. We also use CVSS information for every connected edge weights considering the vulnerability of adjacent nodes and assigned -1 if not connected.


Based on this defined learning heuristic function $f(n)$ over given networks, adaptive tree search should be conducted in terms of highest $f(n)$ value consideration~\cite{9099954} to select the optimal path which makes more effective and damaging attack.

\subsection{MC-based Tree Search for Time-Varying Networks}\label{sec:algo2}
After computing the optimal attack COA paths, our proposed spatio-temporal algorithm additionally conducts MC-based tree search for taking care of time-varying situations over networks~\cite{6145622}. By using the MC-based tree search, the proposed algorithm can identify whether the nearby solutions can be more valuable in terms of attack COA in the scalable and time-varying networks.

\section{Results with Two Example Networks}\label{sec:results}

\begin{figure}[t]
    \begin{center}
        \includegraphics[width=2in]{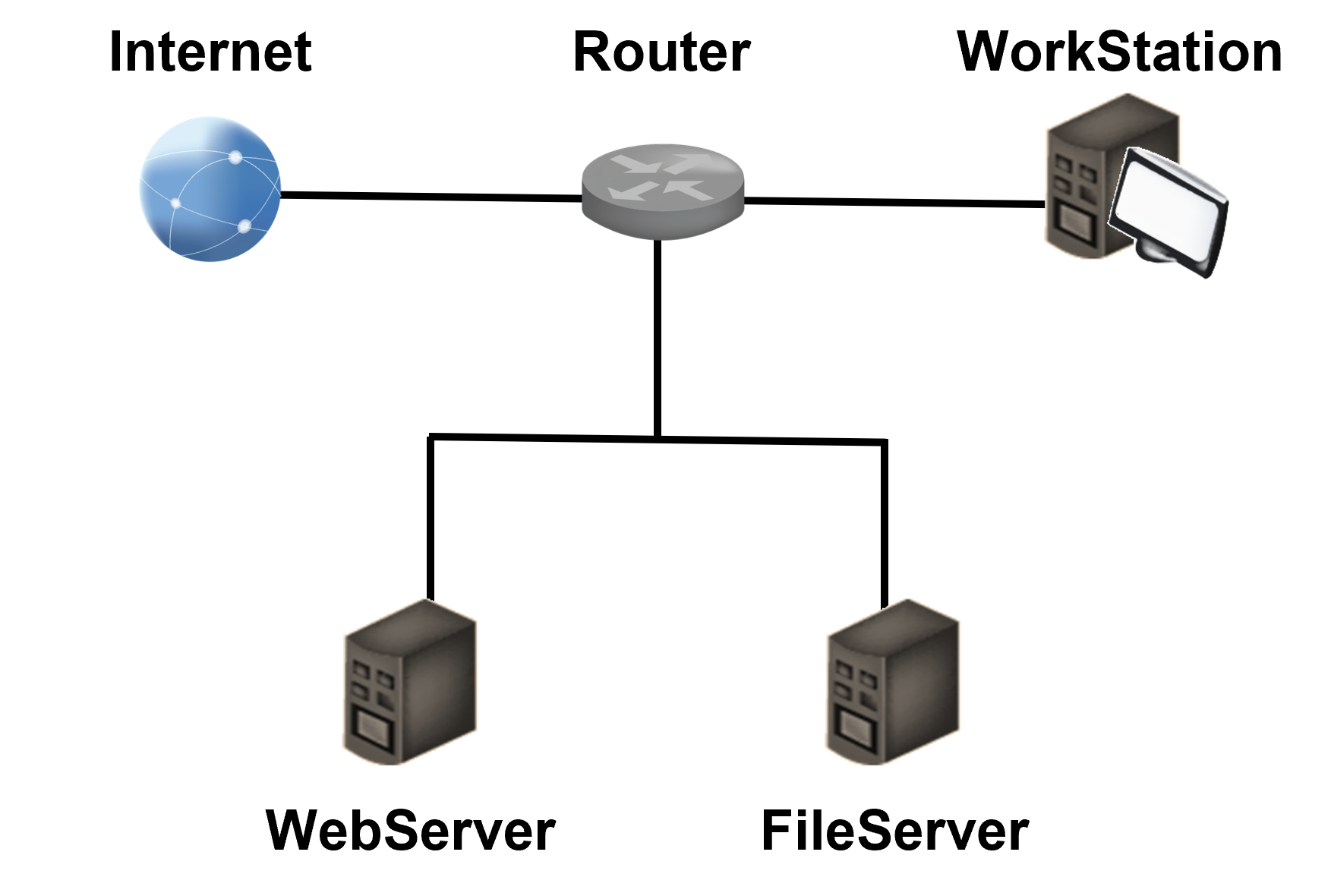}
    \end{center}
    \caption{Network Topology (Example 1)}
    \label{fig1:topology}
\end{figure}

\begin{figure}[t]
    \begin{center}
        \includegraphics[width=3in]{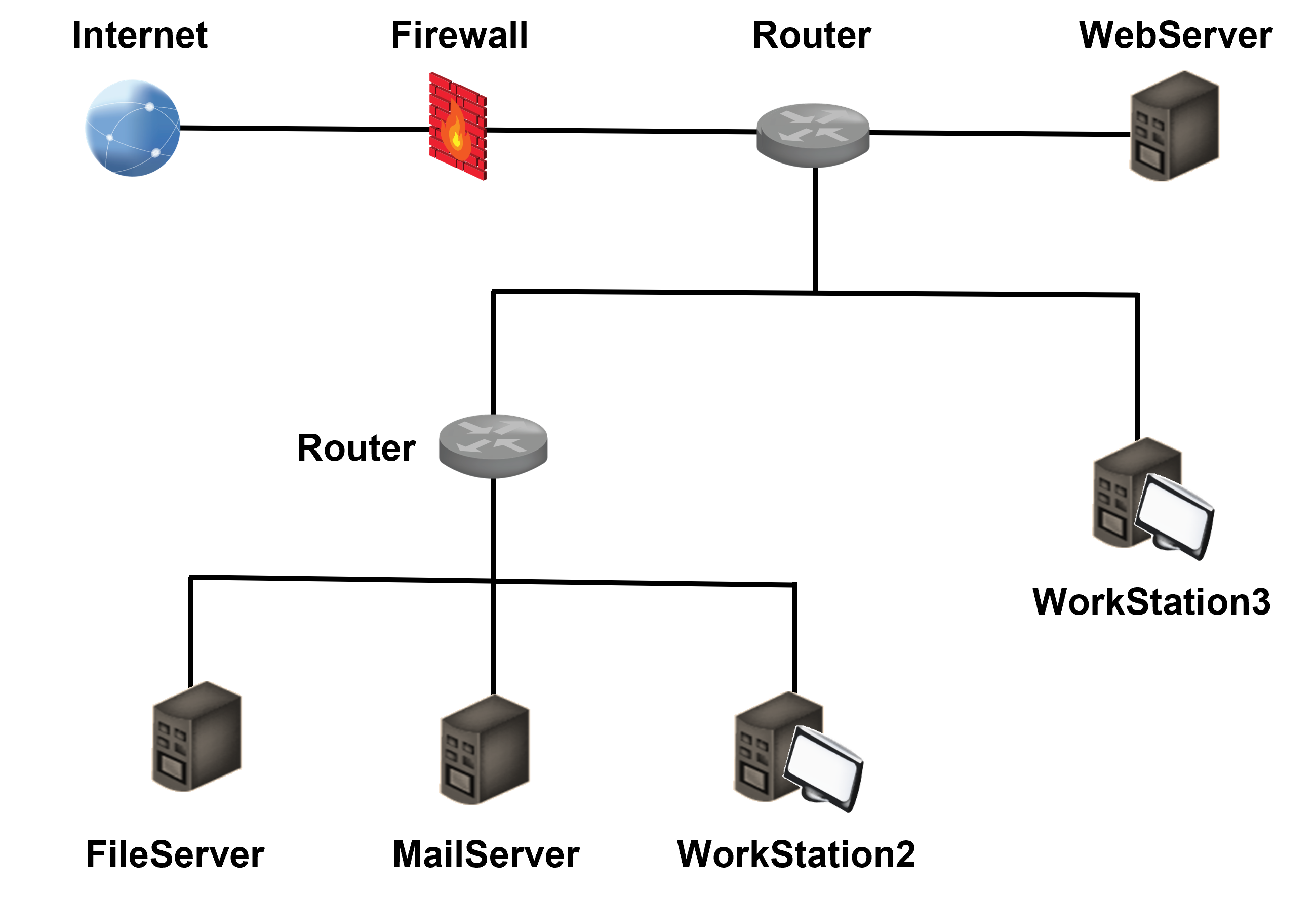}
    \end{center}
    \caption{Network Topology (Example 2)}
    \label{fig2:topology}
\end{figure}

In this paper, two types of networks are considered for evaluating the proposed spatio-temporal attack COA search optimization learning algorithm where the network topology illustrations are in Fig. \ref{fig1:topology} and Fig.~\ref{fig2:topology}.

In this paper, a multi-host multi-stage vulnerability analysis tool (MULVAL) which is one of logic-based security analyzers is used~\cite{mulval}. Note that this is one of widely utilized extendable attack graph analysis tools. This MULVAL generates attack graphs depending on network topology such as Fig.~\ref{fig1:topology} and Fig.~\ref{fig2:topology}. 
As a result, an attack graph is generated. Attack graph is a structured representation of the vulnerabilities present in the system and their interactions with each other. The attack graph is a graph created by the technique of predicting the path for an attacker to break into the system. The attack graph expresses all connections within the network in the form of a graph.It also expresses all possible attack paths based on the Shodan data used in Mulval, vulnerability information such as CVSS Score, and environmental information on the network such as IDS(Intrusion detection system) rules and firewalls.

\begin{figure*}[t]\centering
    \includegraphics[width=7in]{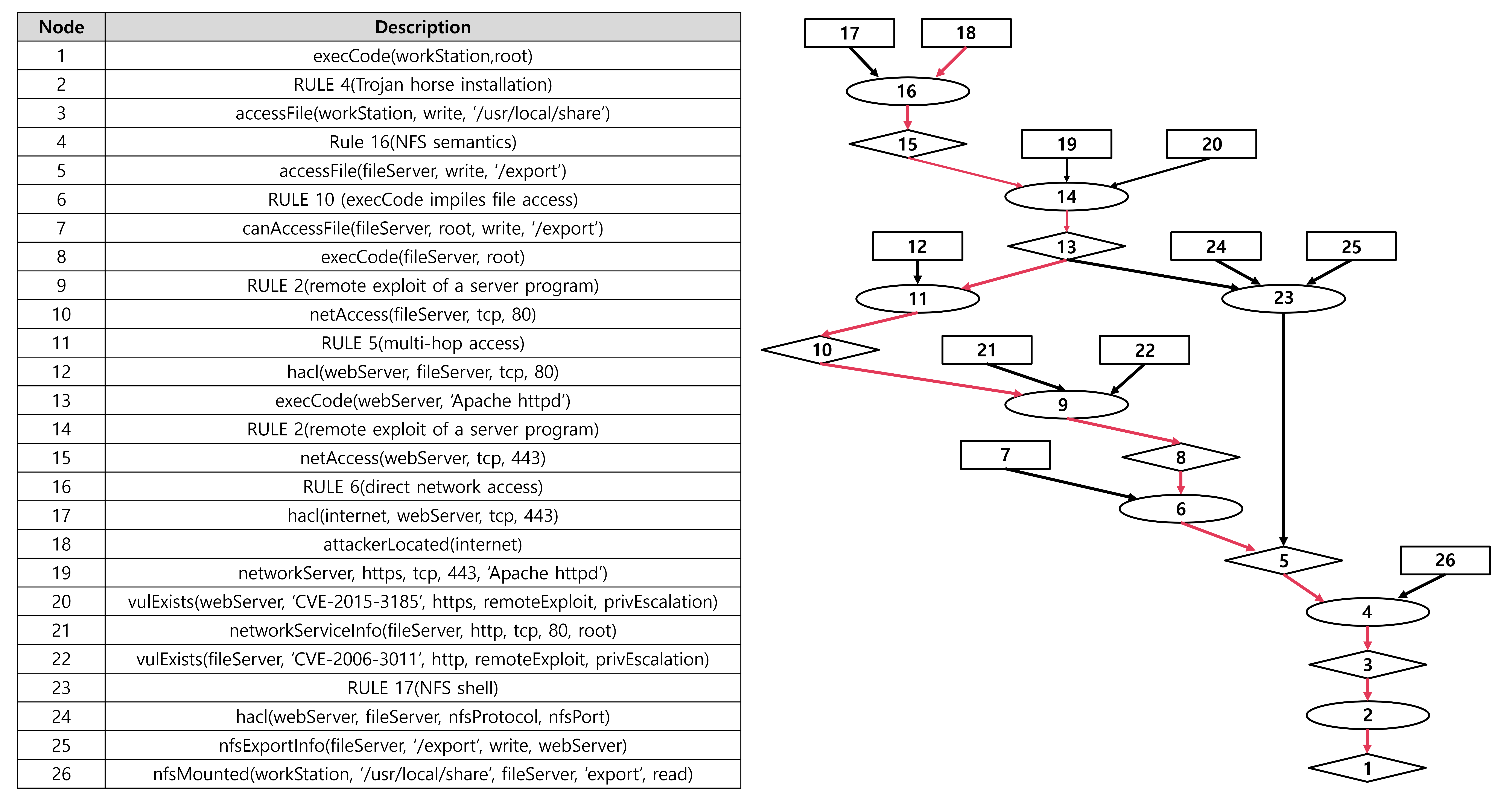}
    \caption{Optimal Attack COA Paths Network Topology (Example 1)}
    \label{fig3:Optimal path of network topology scenario 1 }
\end{figure*}

\begin{figure*}[t]\centering
    \includegraphics[width=7in]{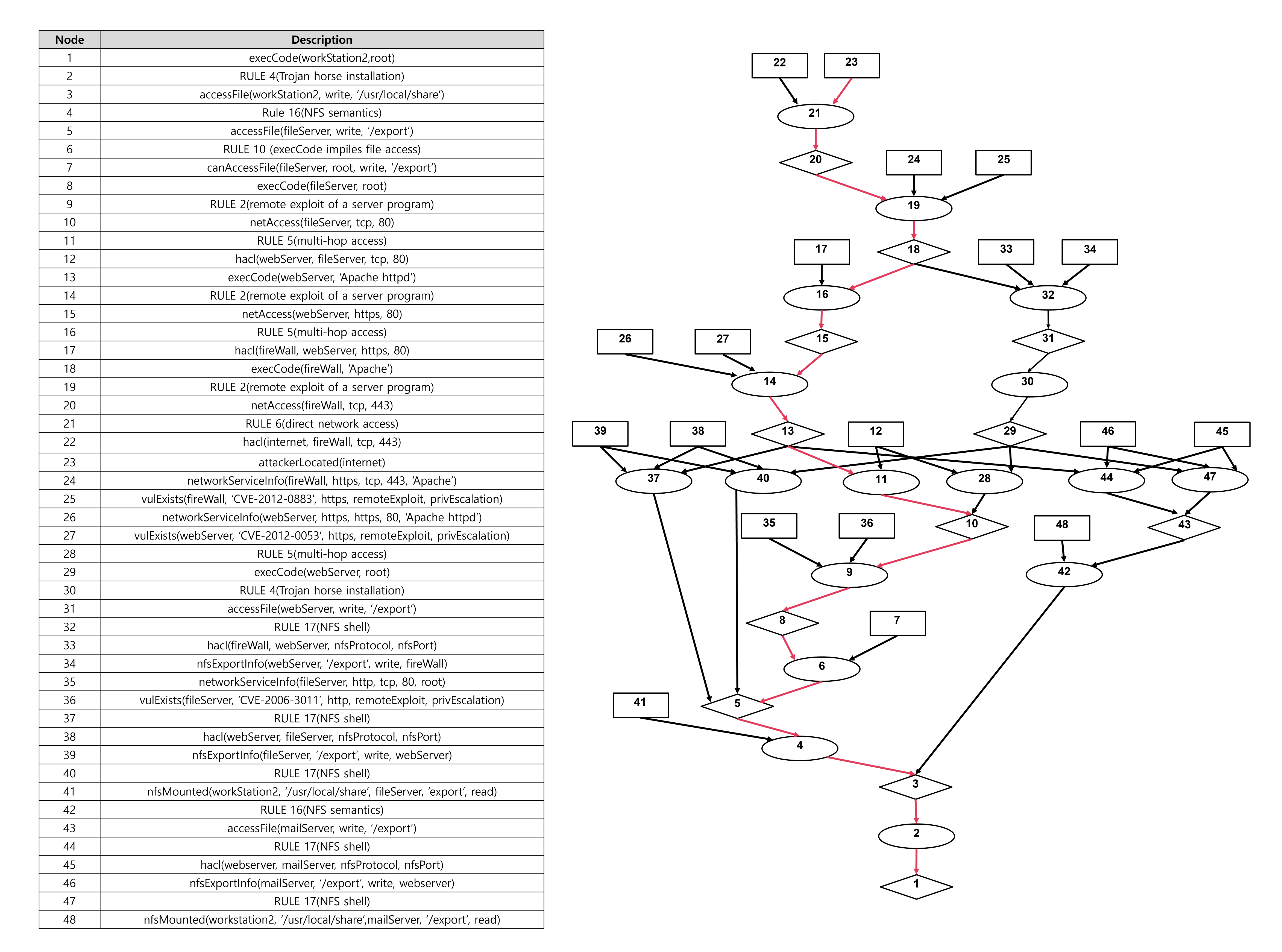}
    \caption{Optimal Attack COA Paths Network Topology (Example 2)}
    \label{fig4:Optimal path of network topology scenario 2}
\end{figure*}

Fig.~\ref{fig3:Optimal path of network topology scenario 1 } and Fig.~\ref{fig4:Optimal path of network topology scenario 2} are the attack graphs generated from the network topology of Fig.~\ref{fig1:topology} and Fig.~\ref{fig2:topology}, respectively. In the Tables of Fig.~\ref{fig3:Optimal path of network topology scenario 1 } and Fig.~\ref{fig4:Optimal path of network topology scenario 2}, various vulnerability information is described, e.g., network connection information, ids rule, and etc. In addition, it can be confirmed that the attack graph is created using the information in the Tables. 
There are various attack paths in the attack graphs, and it is not known which of these paths is the most optimal attack paths. Therefore, finding the optimal attack paths in the attack graphs is critical.

As a result, we confirm that the optimal attack paths can be found using our proposed spatio-temporal attck COA optimization in scalable and time-varying networks those are abstracted as attack graphs. As shown in the red arrows in Fig.~\ref{fig3:Optimal path of network topology scenario 1 }, the optimal attack path is 18 → 16 → 15 → 14 → 13 → 11 → 10 → 9 → 8 → 6 → 5 → 4 → 3 → 2 → 1. Likewise, we can also confirm that the optimal attack COA path in Fig.~\ref{fig4:Optimal path of network topology scenario 2} is as follows: 23 → 21 → 20 → 19 → 18 → 16 → 15 → 14 → 13 → 11 → 10 → 9 → 8 → 6 → 5 → 4 → 3 → 2 → 1, using our proposed spatio-temporal algorithm.

\section{Conclusions and Future Work}\label{sec:conclusions}
In this paper, a novel spatio-temporal attack COA learning algorithm is proposed and evaluated in scalable and time-varying networks. In order to deal with scalable networks, intelligent and adaptive search-based techniques are utilized for designing and implementing our proposed spatial algorithm with appropriate heuristic function definition. In addition, in orrder to deal with time-varying network behaviors, Monte-Carlo (MC) tree search based adaptation is also additionally considered. Therefore, our proposed spatio-temporal attack COA learning algorithm is designed. Furthermore, we confirm that our proposed algorithm works well in two different types of networks. 

As future research directions, our proposed spatio-temporal attack COA learning algorithm should be compared with the other well-known attack COA search methods. Furthermore, considering partially observable networks is also required for more practical applications.

\section*{Acknowledgment}
This work was supported by the Agency for Defense Development under the contract UI210009XD. J. Kim is a corresponding author of this paper.



%



\bibliographystyle{IEEEtran}

\end{document}